# Impact of Data-Driven Eddy Parameterization on Climate State in an Idealized Coupled CESM Model

Jia-Rui Shi[1], Pavel Perezhogin[1], Laure Zanna[1], and Alistair Adcroft[2]

[1]*Courant Institute School of Mathematics, Computing, and Data Science, New York University, New York, NY, USA*

[2]*Atmospheric and Oceanic Sciences Program, Princeton University, Princeton, USA*

\* Corresponding author: Jia-Rui Shi (jia-rui.shi@nyu.edu)

## Abstract

Mesoscale eddies remain poorly represented in most climate models, motivating the use of parameterizations to account for their dynamical effects on the coupled system. In this study, we implement a data-driven eddy parameterization based on Zanna and Bolton (2020; ZB20) in an idealized, fully coupled CESM configuration and assess its influence on the mean climate state. When applied within an eddy-permitting ocean model (MOM6) embedded in the coupled configuration, the ZB20 eddy momentum parameterization, which features upgradient (backscatter) momentum flux, energizes mesoscale eddies and strengthens poleward ocean heat transport. The response is particularly strong in the Southern Hemisphere, where the open circumpolar channel sustains vigorous eddy activity and is sensitive to the parameterization, further leading to a marked hemispheric asymmetry. The oceanic meridional overturning circulation also intensifies around 60°S. The resulting ocean adjustments produce a coherent dipolar temperature pattern, with cooling in mid-latitudes and warming at high latitudes, driven primarily by anomalous meridional heat transport rather than local surface fluxes, shown using a regional heat-budget analysis. The atmosphere, in turn, exhibits a compensating reduction in meridional heat transport and an equatorward shift of the mid-latitude jet, associated with the mid-latitude surface cooling and changes in the meridional temperature gradient. Together, these results highlight how a data-driven eddy momentum parameterization can affect large-scale circulation and the mean climate state, providing a reference for understanding its impacts in more comprehensive climate models.

**Plain Language Summary:**

Mesoscale eddies help move ocean water masses, tracers, and heat, but most climate models cannot resolve them and rely on parameterizations. We test a data-driven eddy momentum parameterization in an idealized, fully coupled climate model and assess changes in the mean climate state. The parameterization increases eddy activity and strengthens poleward ocean heat transport, with the strongest response in the Southern Hemisphere where an open circumpolar channel supports vigorous eddies. These ocean changes produce cooling in mid-latitudes and warming at high latitudes. A regional heat-budget analysis shows this pattern mainly results from changes in meridional ocean heat transport, not local surface heat fluxes. The atmosphere responds by transporting less heat poleward and shifting the mid-latitude jet equatorward.

# 1. Introduction

Oceanic mesoscale eddies, features with horizontal scales of 10 km near the polar regions to 100 km near the equator (Chelton et al. 1998), play a fundamental role in the ocean's general circulation, transport of heat, salt and biogeochemical tracers (Redi 1982; Ferrari and Wunsch 2009; Griffies et al. 2015; Uchida et al. 2017), and further affect our climate (Delworth et al. 2012; Beech et al. 2022; Bian et al. 2023). They emerge primarily from baroclinic instabilities associated with steeply sloping isopycnals, which convert available potential energy into eddy kinetic energy (Rintoul 2018).

Large-scale ocean circulation (e.g. the meridional overturning circulation and basin-scale gyres) can be well resolved in climate models; however, computational constraints limit the achievable horizontal resolution required to fully resolve mesoscale eddies (Hewitt et al. 2020; Christensen and Zanna 2022). Adequately resolving mesoscale eddies requires a grid spacing of at least about two grid intervals per first baroclinic deformation radius (Hallberg 2013). Consequently, the effects of unresolved mesoscale motions on the large-scale circulation must be parameterized. Ocean mesoscale eddy parameterizations, such as Gent and McWilliams (1990), are designed to capture the impacts of unresolved eddies on the mean flow, and are essential for obtaining accurate mean ocean state, climate variability, and future climate response (Hewitt et al. 2020).

Furthermore, at eddy-permitting resolutions, the upscale energy transfer from unresolved motions is not fully represented, leading to an under-energetic eddy field and biases in the mean circulation and tracer transport (Jansen and Held 2014). To mitigate the underestimation of eddy kinetic energy at eddy-permitting resolutions, several studies have introduced kinetic energy backscatter schemes that apply an effective anti-viscosity to represent the upscale transfer of kinetic energy from unresolved motions to the resolved flow (e.g., Jansen and Held 2014; Jansen et al. 2015; Klöwer et al. 2018; Juricke et al. 2019). In parallel, data-driven eddy parameterizations have emerged as an alternative approach to capture backscatter (Zanna and Bolton 2020; Frezat et al. 2022; Guillaumin and Zanna 2021; Guan et al. 2022). For instance, the ZB20 (Zanna and Bolton 2020) eddy parameterization strategy uses high-resolution simulations that explicitly resolve mesoscale eddies as a reference, from which coarse-grained data are derived to diagnose the missing subgrid fluxes. Machine learning models are then trained to infer these fluxes from the coarse-resolution fields and validated on independent datasets to assess their generalization and physical consistency. The implementation of eddy parameterizations that account for backscatter, whether data-driven or not, has been shown to improve the simulated energy distribution and mean flow, and to further reduce climatological biases between models and observations (e.g., Juricke et al. 2019; C. Zhang et al. 2023; Perezhogin et al. 2024).

Previous studies have primarily examined the impact of backscatter schemes in ocean-only or ocean–ice models. However, oceanic and atmospheric circulations are dynamically coupled through air–sea exchanges of momentum, heat, and freshwater (Wunsch 2005; Trenberth and Caron 2001; Stephens et al. 2012). The ocean mixed layer responds to atmospheric forcing while simultaneously feeding back to the atmosphere, influencing large-scale circulation, weather, and climate state and variability. Accordingly, the climatic impacts of oceanic backscatter schemes can be substantially modulated by air–sea interactions, which are not fully captured in uncoupled model configurations. Therefore, implementing such parameterizations within a fully coupled global model provides a critical step toward understanding the climatic impacts of backscatter.

The climate impacts of eddy parameterizations remain difficult to interpret in fully coupled climate models, where responses are intertwined with complex geometry, coupled feedback, and numerous interacting parameterizations. This complexity makes it challenging to attribute changes in large-scale circulation and tracer transport to a specific process such as kinetic energy backscatter. Simplifying the model configuration can reduce this complexity and thereby help isolate and understand key physical processes across different spatial and temporal scales (Farneti and Vallis 2009; Wolfe and Cessi 2010; Abernathey et al. 2013). Previous studies using idealized ocean geometries have demonstrated that such simplified frameworks can still reproduce essential features of the observed climate, including meridional heat transport and large-scale circulation patterns (Enderton and Marshall 2009; Smith et al. 2006; Wu et al. 2021).

In this work, we implement the ZB20 subgrid-scale parameterization into an idealized, coupled global CESM configuration featuring a ridge ocean geometry with an open channel in the Southern Hemisphere, analogous to the Drake Passage (see Section 2.1). The ocean component is run at eddy-permitting resolution, allowing the parameterization to act effectively on the upscale transfer of kinetic energy. We focus on the climatic impacts of the backscatter component, examining its influence on the mean climate state, regional temperature and wind patterns, large-scale circulation, and meridional heat transport in both the ocean and atmosphere. Although the simplified bathymetry precludes direct comparison with observations, this configuration enables a clear diagnosis of the simulated responses and underlying physical processes induced by the parameterization. The insights gained from this framework can inform the development and tuning of eddy parameterizations in more realistic climate models.

The paper is organized as follows. Section 2 describes the model configuration, the subgrid momentum parameterization, and the experimental design. Section 3 presents the mean-state responses of the ridge configuration under the parameterization. Section 4 shows the results of sensitivity tests. Section 5 summarizes the main findings, discusses the physical mechanisms and broader implications of the backscatter scheme in the coupled climate system, and outlines directions for future work.

## 2. Data and Methods

### 2.1 Configurations of the Idealized Coupled Model

The idealized experiments are conducted using the Community Earth System Model (CESM; Danabasoglu et al. 2020; Hurrell et al. 2013), a comprehensive Earth system modeling framework widely used for both process-level investigations and international climate assessments (Eyring et al. 2016). CESM offers a flexible architecture that enables components of varying complexity to be coupled together, making it well-suited for systematically isolating physical mechanisms in simplified configurations. In this study, we adopt a fully coupled setup in which the ocean, atmosphere, land, and sea-ice components interact dynamically based on the coupled Aqua and Ridge planets framework (Wu et al. 2021). Our configuration differs from Wu et al. (2021) in horizontal resolution and the details of the ridge/bathymetry setup (see below). Real-world geographic complexity (continents, bathymetry, and detailed coastlines) is removed. This idealization allows us to isolate the dynamical impacts of the backscatter-enabled eddy parameterization in a controlled coupled setting (e.g. Enderton & Marshall, 2009).

The ocean model is based on the Modular Ocean Model version 6 (MOM6; Adcroft et al. 2019), which has recently become the ocean dynamical core of forthcoming CESM3 releases. MOM6 has been widely adopted in both idealized and comprehensive ocean–climate applications, reflecting its flexible vertical coordinate framework, numerical robustness, and suitability for a broad range of experimental designs. The ocean is configured on a 0.25° × 0.25° horizontal grid, placing the model in the eddy-permitting regime and allowing the ZB20 eddy momentum parameterization to operate on the upscale kinetic-energy transfer. The model uses the stretched geopotential $Z^*$ coordinate, which closely resembles a standard z-level but is formulated to accommodate sizable free-surface variations (Adcroft et al. 2019). The ocean has a maximum depth of 4400 m, discretized into 15 vertical layers with a surface layer thickness of 20 m and a bottom layer thickness of 500 m. In this study, we disable the Gent–McWilliams (GM) parameterization to isolate the impact of the data-driven subgrid momentum fluxes. We employ the ZB20 eddy momentum parameterization (see Section 2.2). To ensure numerical stability, the downgradient component of the momentum flux is represented by a biharmonic Smagorinsky parameterization (Griffies and Hallberg 2000; Perezhogin et al. 2024), while ZB20 provides the backscatter (upgradient) contribution. The non-dimensional Smagorinsky coefficient is 0.06.

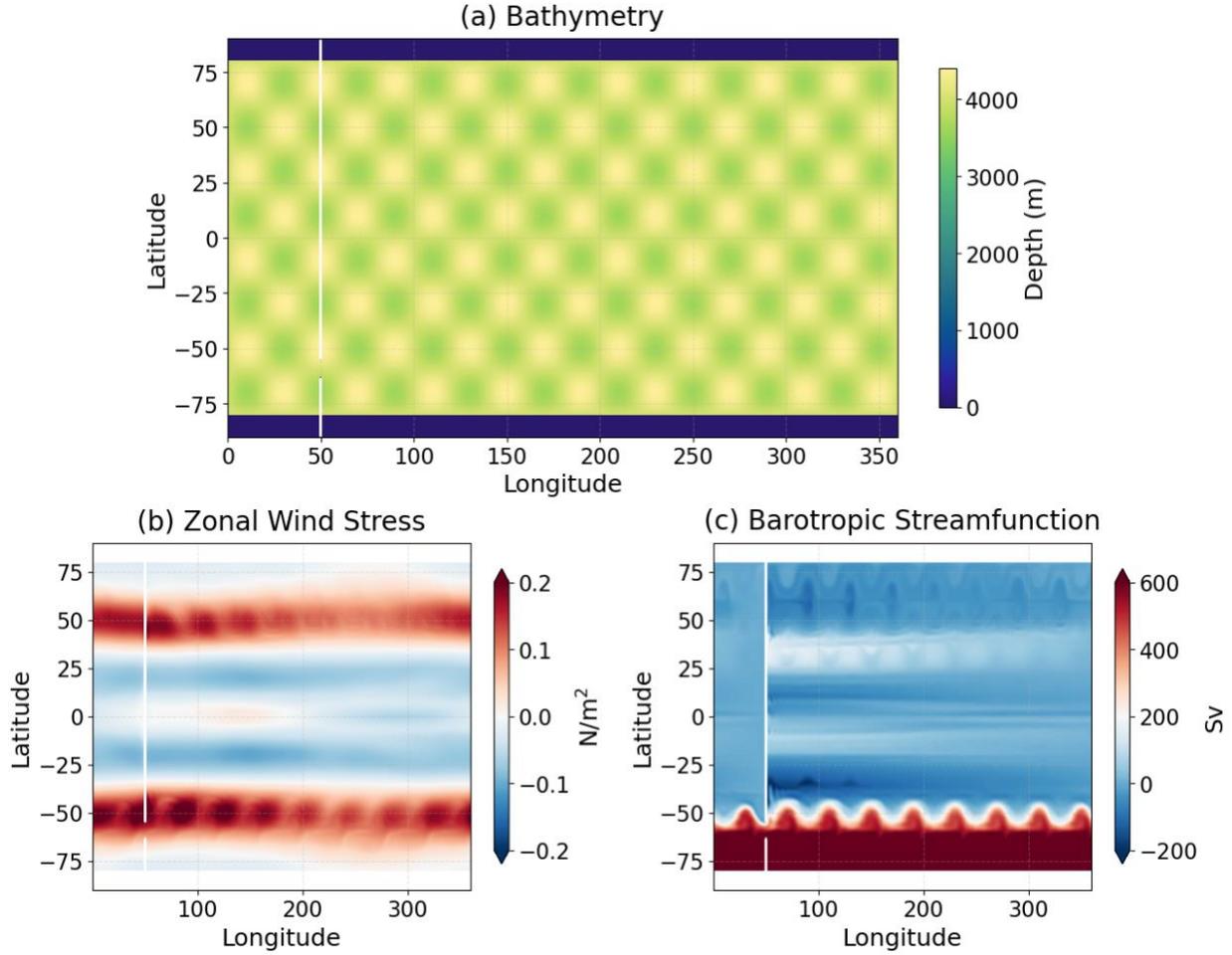

**Figure 1**. (a) Idealized bathymetry and geometry used in the coupled CESM model, (b) 40-year climatology of surface zonal wind stress (in N/m²), and (c) barotropic streamfunction (in Sv). For bathymetry, a meridional ridge from pole to pole with an open circumpolar channel in the Southern Hemisphere. Polar regions poleward of 80°N/S are land. Bottom topography follows a mean depth of 4000 m with a ±400 m sinusoidal amplitude.

A highly simplified ocean geometry is used to minimize geographic complexity while retaining essential large-scale dynamical features. The configuration follows a "ridge planet" design (e.g., Enderton and Marshall 2009; Smith et al. 2006; Wu et al. 2021): a single meridional land barrier extends from pole to pole, and an open circumpolar channel in the Southern Hemisphere (Figure 1a). This channel acts as an idealized analog of the Drake Passage and supports strong circumpolar flow that interacts with mesoscale eddies and the energy backscatter scheme. The channel is open from the surface to the bottom, consistent with the configuration of Enderton and Marshall (2009). The polar regions poleward of 80°N/S are represented as land, and all land points have zero orography. The bottom topography has a mean depth of 4000 m, superimposed on a large-scale sinusoidal perturbation with an amplitude of ±400 m. This perturbation provides bottom form drag and enhances numerical stability, following previous idealized studies (e.g. Wu et al. 2021). Figure 1 also shows the 40-year climatology of surface zonal wind stress (Figure 1b) and the barotropic streamfunction (Figure 1c) as illustrative examples.

The atmosphere is simulated using the Community Atmosphere Model version 4 (CAM4; Neale et al. 2013), configured with the finite-volume dynamical core on a regular latitude–longitude grid. The atmospheric model is run at a nominal 1° horizontal resolution with 26 vertical levels in a hybrid sigma–pressure coordinate system. $CO_2$ and other external forcings are fixed at preindustrial values. The model retains a full diurnal cycle, and an idealized seasonal cycle is imposed by prescribing an orbital obliquity of 23.3°. The sea-ice component is the Community Ice CodE (CICE; Bailey et al. 2018). Under this coupled configuration, the climate is sufficiently warm that no sea ice forms, even though the sea-ice model remains active. The land component is represented by the Community Land Model version 5 (CLM5; Lawrence et al. 2019), which is configured at the same horizontal resolution as the atmosphere. Coupling among all model components is handled through the Common Infrastructure for Modeling the Earth (CIME; Danabasoglu et al., 2020). All the above atmospheric, sea-ice, land and coupling configurations are identical across the Control and parameterized experiments (see Section 2.3).

## 2.2 Data-driven Eddy Parameterization

We use the parameterization of ocean eddy momentum forcing of Zanna and Bolton (2020) (ZB20), implemented and adjusted for the MOM6 by Perezhogin et al. (2024).

The eddy parameterization modifies the momentum equation as follows:

$$\partial_t \boldsymbol{u} = \cdots + \nabla \cdot \mathbf{T} \qquad \text{(Eq. 1)}$$

where $\boldsymbol{u}$ is the horizontal velocity, $\nabla \cdot$ is the divergence operator $\nabla = (\partial_x, \partial_y)$, the ellipsis denotes the standard momentum terms (e.g., advection, Coriolis force, and pressure gradient), and

$$\mathbf{T} = \begin{bmatrix} T_{xx} & T_{xy} \\ T_{xy} & T_{yy} \end{bmatrix} \qquad \text{(Eq. 2)}$$

is the symmetric horizontal stress tensor. The ZB20 parameterization predicts the stress tensor as follows:

$$\mathbf{T}_{ZB20}(\zeta, D, \widetilde{D}) = \gamma \Delta^2 \begin{bmatrix} \zeta D & -\zeta \widetilde{D} \\ -\zeta \widetilde{D} & -\zeta D \end{bmatrix} - \frac{\gamma \Delta^2}{2}(\zeta^2 + D^2 + \widetilde{D}^2)\begin{bmatrix} 1 & 0 \\ 0 & 1 \end{bmatrix}, \qquad \text{(Eq. 3)}$$

where $\Delta$ is the grid spacing, $\gamma = O(1)$ is the nondimensional tunable scaling coefficient, and $\zeta = \partial_x v - \partial_y u$ is the relative vorticity, $D = \partial_x v + \partial_y u$ is the shearing deformation, $\widetilde{D} = \partial_x u - \partial_y v$ is the stretching deformation.

We employ a filtering scheme to enhance numerical stability and extract the backscattering effect from the original ZB20 parameterization. Perezhogin et al. (2024) introduced a filtering scheme, referred to as ZB20-Reynolds, which extracts the component of the Germano (1986) decomposition responsible for the strongest backscatter – the Reynolds stress (Perezhogin and Glazunov 2023). The filtering scheme applies a high-pass filtering, denoted as $(\cdot)'$, to the velocity gradients and low-pass filtering, denoted as $G$, to the stresses predicted by the ZB20 parameterization such that

$$\mathbf{T} = G(\mathbf{T}_{ZB20}(\zeta', D', \widetilde{D}')). \quad \text{(Eq. 4)}$$

Therefore, this filtering scheme alters the ZB20 parameterization to predict stronger backscatter, that is, higher $\mathbf{u} \cdot (\nabla \cdot \mathbf{T}) > 0$ on average, given the numerical stability constraints (Perezhogin et al. 2024).

**2.3 Experiments Setup**

We conduct a suite of experiments to assess the climatic impacts of the ZB20 Reynolds subgrid momentum parameterization. The baseline Control experiment excludes the ZB20 parameterization but retains the biharmonic Smagorinsky viscosity. To obtain a dynamically balanced initial state, we first integrate a standard-resolution (1° × 1°) fully coupled configuration for 125 years using the same atmospheric, land, and sea-ice settings described in Section 2.1. The final state of this integration is remapped to the eddy-permitting 0.25° × 0.25° MOM6 grid to initialize the high-resolution simulations. The remapped configuration is then spun up for an additional 250 years to reduce initial transients and allow the ocean to adjust toward the finer resolution.

From this spun-up state, we branch two main experiments: (1) an unparameterized Control run, and (2) a ZB20-Reynolds run (hereafter ZB20) that activates the data-driven eddy parameterization with a scaling coefficient of 2.0. Both experiments are integrated for 40 years, and differences between them are interpreted as the climatic response to the ZB20 parameterization. While full equilibration of the deep ocean may require substantially longer integrations, the primary objective here is to quantify the decadal-scale mean-state response and its underlying mechanisms associated with kinetic energy backscatter. A 40-year integration also provides sufficient sampling to reduce the influence of internal variability on the estimated mean response.

In addition, we conduct a set of sensitivity experiments by varying the ZB20 scaling coefficient ($\gamma$; see Section 2.2). Additional sensitivity experiments are performed using coefficients of 1.5, 2.5, and 3.0, with each run integrated for 10 years. Longer sensitivity integrations would further reduce sampling uncertainty but are computationally expensive in a fully coupled, eddy-permitting configuration. In practice, the 10-year sensitivity runs provide a clear and consistent mean-state response across different values of γ, allowing us to assess the dependence on backscatter amplitude.

## 3. Results

### 3.1 Response of Kinetic Energy

Figure 2 shows the impact of the data-driven eddy parameterization on the oceanic kinetic energy (KE). The inclusion of the backscatter scheme leads to an increase in total KE across most ocean basins, indicating enhanced mesoscale activity. The global KE increases approximately 13% (with a scaling coefficient γ=2.0) relative to the control run. The KE enhancement is spatially heterogeneous, with the largest increases found in the mid-latitude regions of both hemispheres, particularly between 30° and 60° and along the western boundary system. The existence of the "Drake Passage," an open channel in the Southern Hemisphere mid-latitudes, allows strong circumpolar flow, baroclinic instability, and promotes an eddy-rich regime in the "Southern Ocean". Consequently, the KE response to backscatter is more pronounced in the Southern Hemisphere (Figure 2c), consistent with the broader and more energetic zonal current systems and the dominance of the "Antarctic Circumpolar Current".

At lower latitudes, the KE increase is weaker, reflecting modest intensification of the equatorial currents. Notably, the kinetic energy maximum remains collocated with regions of strong mean flow and baroclinic instability, suggesting that the backscatter scheme primarily amplifies dynamically active regions rather than generating new mesoscale activity elsewhere. Vertically, climatological KE peaks within the upper 100–1000 m, and the backscatter-induced increase is likewise most pronounced in this layer (Figure 2d).

To further understand these energetic adjustments, the total KE change is decomposed into the contributions from mean KE (MKE) and eddy KE (EKE), respectively. This decomposition allows separation between changes in the time-mean circulation and those associated with transient eddies. The change in total KE is dominated by EKE (Figure 2e), indicating that backscatter primarily energizes transient eddies. Consistent with the total KE pattern, EKE increases are strongest in the mid-latitude bands of both hemispheres. Furthermore, for total KE and EKE, the Southern Hemisphere shows a broad, basin-spanning enhancement, whereas the Northern Hemisphere response is narrower and spike-like along major jets and boundary currents (Figure 2e).

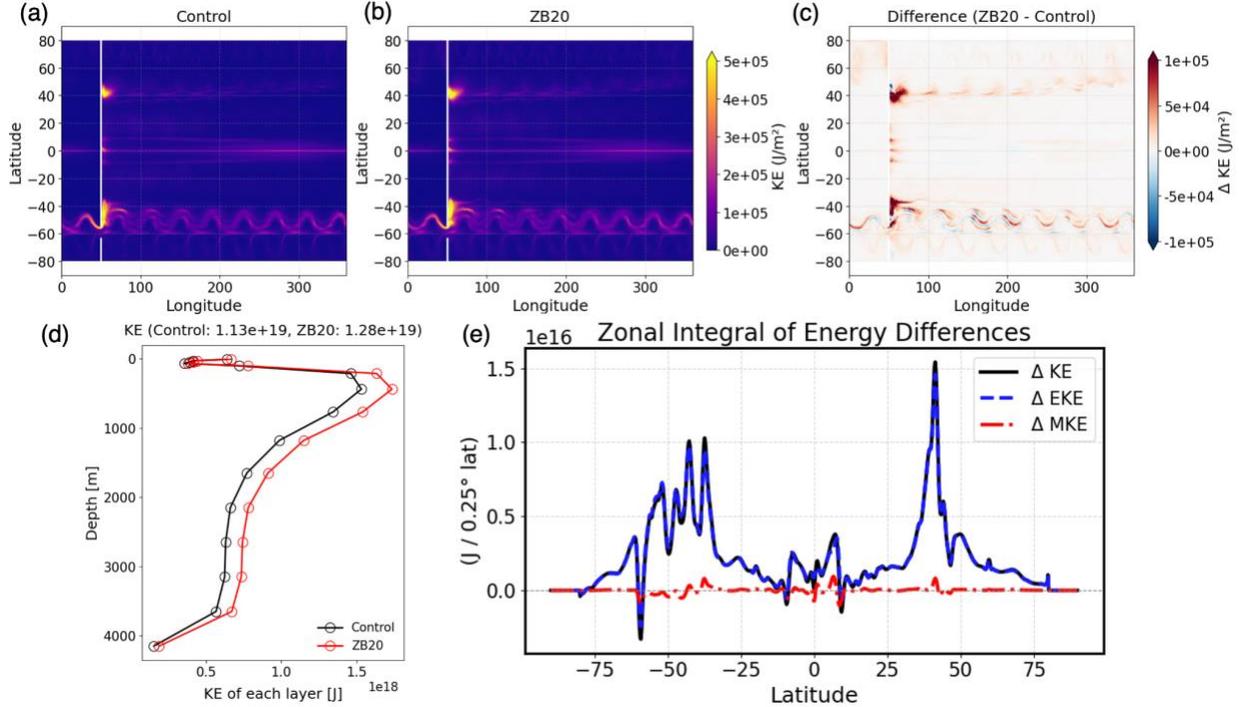

**Figure 2.** Climatological ocean kinetic energy (KE) and its response to the data-driven eddy parameterization (e.g. ZB20 minus Control) averaged over 40 years. Panels show (a) vertically integrated KE in the Control simulation (J m$^{-2}$), (b) KE in the ZB20 simulation (J m$^{-2}$), (c) KE difference (ZB20 − Control; J m$^{-2}$), (d) vertical structure of layer-integrated KE (in J) from Control (black) and ZB20 (red), and (e) decomposition of KE response (black curve) into mean KE (MKE; red curve) and eddy KE (EKE; blue curve) components, integrated within each 0.25° latitude band.

Figure 3 shows the surface EKE wavenumber spectrum averaged over 80°S–40°S. The backscatter scheme produces a clear enhancement of eddy energy at scales between roughly 10 km and 25 km, corresponding to the local first baroclinic deformation radius. The spectral shape, however, remains nearly unchanged between the Control and ZB20 simulations: no systematic shift toward larger or smaller eddy scales. This indicates that the subgrid momentum parameterization primarily increases the eddy velocity scale, rather than modifying the eddy length scale. In the framework of mixing-length theory, the isopycnal diffusivity scales with the product of the eddy velocity and eddy length scales. This implies stronger lateral mixing along isopycnals due to enhanced EKE.

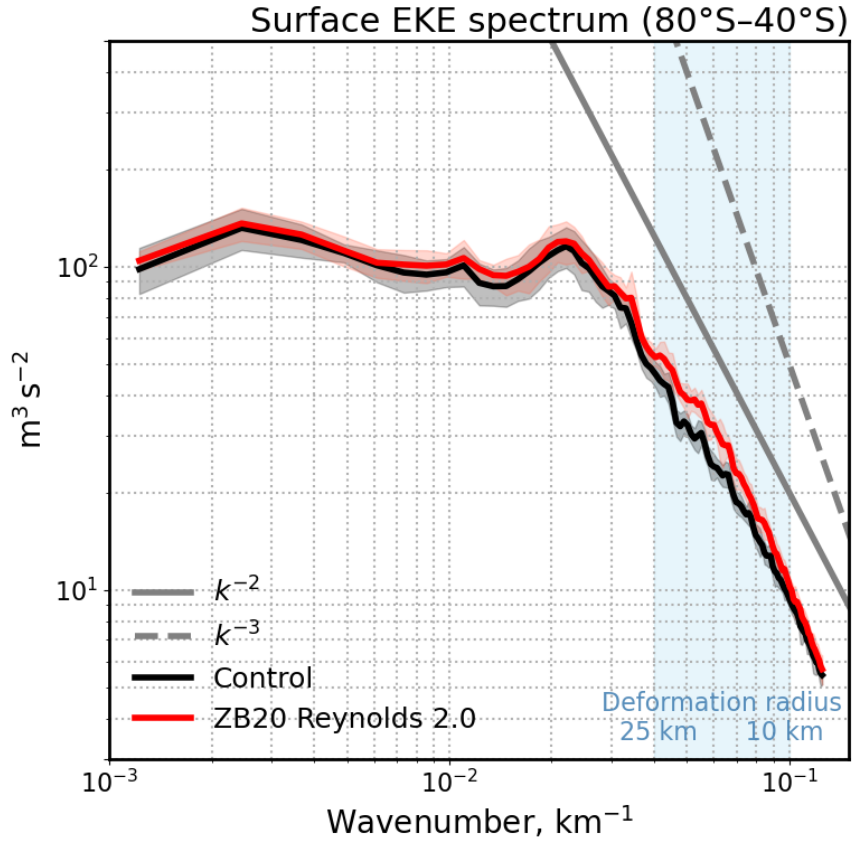

**Figure 3.** Surface eddy kinetic energy (EKE) zonal wavenumber spectra averaged over 80°S–40°S for Control (black) and ZB20 (red) simulations. Solid curves show the spectra averaged over 40 years, and shading indicates ±1 standard deviation across the annual spectra. Gray reference lines denote the k$^{-2}$ (solid) and k$^{−3}$ (dashed) slopes.

## 3.2 Response of Sea Surface Temperature

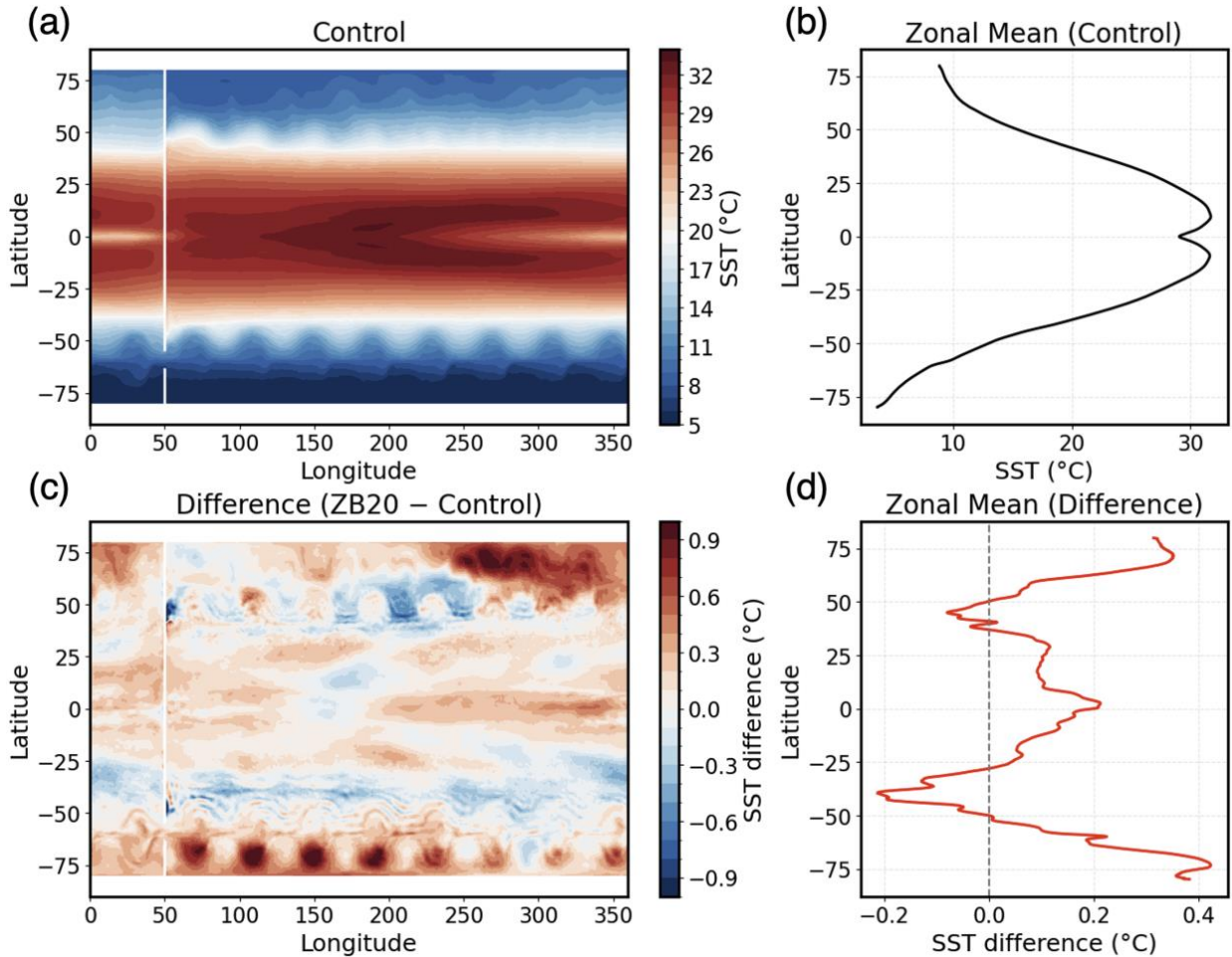

**Figure 4.** Sea surface temperature (SST) mean state and response to the ZB20 parameterization for the last 20 years of simulation. Panels show (a) the climatological SST map and (b) its zonal mean for the Control simulation, and (c,d) the corresponding SST anomalies induced by the ZB20 parameterization (ZB20 − Control).

Figure 4a,b show the simulated mean sea surface temperature (SST) in the Control run. The large-scale structure exhibits warm tropical waters exceeding 28 °C and progressively cooler temperatures toward the poles. The polar regions exhibit temperatures around 5 °C, with relatively higher values in the Northern Hemisphere than in the Southern Hemisphere. The elevated temperatures inhibit sea ice formation in both polar regions. In the mid-latitudes, the mean SST pattern and ocean circulation is largely steered by the idealized bathymetry. Consistent with Wu et al. (2021), the equatorial region exhibits relatively low SST compared with the regions off the equator, reflecting the upwelling of cold subsurface waters along the equator and the presence of a strong cold tongue in the eastern basins. It is analogous to the Pacific cold tongue in the real

ocean which is maintained by equatorial wind stress and ocean–atmosphere coupling. The model thus captures essential zonal and meridional SST gradients and key dynamical features of the coupled system, despite its idealized configuration.

Figure 4c,d illustrate the SST response to the data-driven eddy parameterization, highlighting the backscatter effect on the coupled climate state. A distinct mid-latitude cooling is evident in both hemispheres, with temperature decreases of approximately 0.2 °C between 30°S and 60°S latitude. In contrast, the higher-latitude oceans (poleward of ~60°S) exhibit a notable warming of 0.4°C. This dipolar pattern, of mid-latitude surface cooling accompanied by high-latitude warming, suggests a net poleward shift of ocean heat transport associated with enhanced eddy activity under the backscatter scheme (Figure 2). The SST changes can feed back onto the atmosphere, modifying the surface heat fluxes and shifting the meridional heat flux convergence, thereby altering the large-scale atmospheric circulation. These coupled adjustments will be further examined in Section 3.3-3.5.

### 3.3 Atmospheric Responses

Figure 5 shows the zonal-mean atmospheric responses to the data-driven eddy backscatter scheme. The westerly wind in the Southern Hemisphere exhibits a distinct dipolar anomaly, characterized by weakening on the poleward flank and strengthening on the equatorward flank. This pattern indicates an equatorward shift of the mid-latitude jet, consistent with the SST cooling and altered meridional temperature contrast described in Section 3.2. The equatorward displacement of the jet also implies a shift of the baroclinic zones and storm tracks, reflecting adjustments of large-scale atmospheric circulation in response to the modified surface thermal forcing.

The zonal-mean atmospheric temperature response closely mirrors the oceanic SST pattern: a mid-latitude cooling accompanied by high-latitude warming, particularly in the Southern Hemisphere. This vertical and meridional temperature redistribution leads to changes in the meridional temperature gradient, which, through the thermal wind relation, induces the modifications in the zonal wind structure.

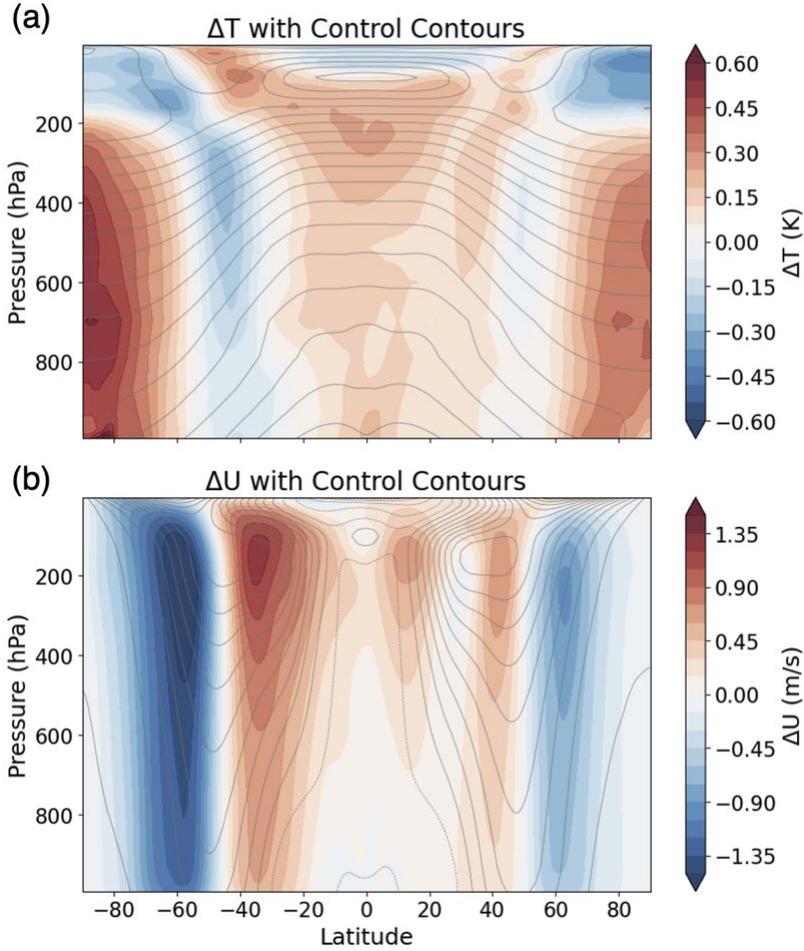

**Figure 5.** Zonal-mean atmospheric responses to the ZB20 parameterization averaged over the last 20 years of simulation. Shading shows anomalies (ZB20 - Control) in (a) atmospheric temperature and (b) zonal wind, and contours show the corresponding climatological fields from Control. Solid (dashed) contours in (b) denote the westerly (easterly) winds.

## 3.4 Subsurface Responses

Figure 6 presents the subsurface temperature response to the data-driven eddy backscatter scheme. Consistent with the surface pattern, the most prominent feature is a mid-latitude subsurface cooling accompanied by high-latitude subsurface warming. The cooling signal is concentrated within the main thermocline (Figure 6a), where temperature anomalies reach amplitudes substantially larger than those at the sea surface. The cooling signal largely aligns with regions of strong baroclinic instability and mean isopycnal slopes, consistent with eddy-driven isopycnal flattening and strengthened along-isopycnal mixing. In contrast, subpolar regions exhibit a deep-reaching warming that extends to the bottom, associated with deep convection. This dipolar temperature response suggests that the backscatter-induced resolved bolus velocity and enhancement of mesoscale eddy activity (Figure 2) intensify the heat exchange between mid-latitudes and higher-latitudes, promoting more efficient ocean poleward heat transport (see Section 3.5). The potential density response exhibits a similarly organized dipolar structure: density increases in the mid-latitudes and decreases in the subpolar region (Figure 6b). The stratification strengthens markedly at high latitudes (Figure 6c), while the mixed layer shoals (dashed curves in Figure 6b), indicating a more stable upper ocean that resists vertical convection. These features are physically linked to the additional surface and subsurface warming in subpolar regions, which enhances vertical temperature gradients and suppresses deep mixing.

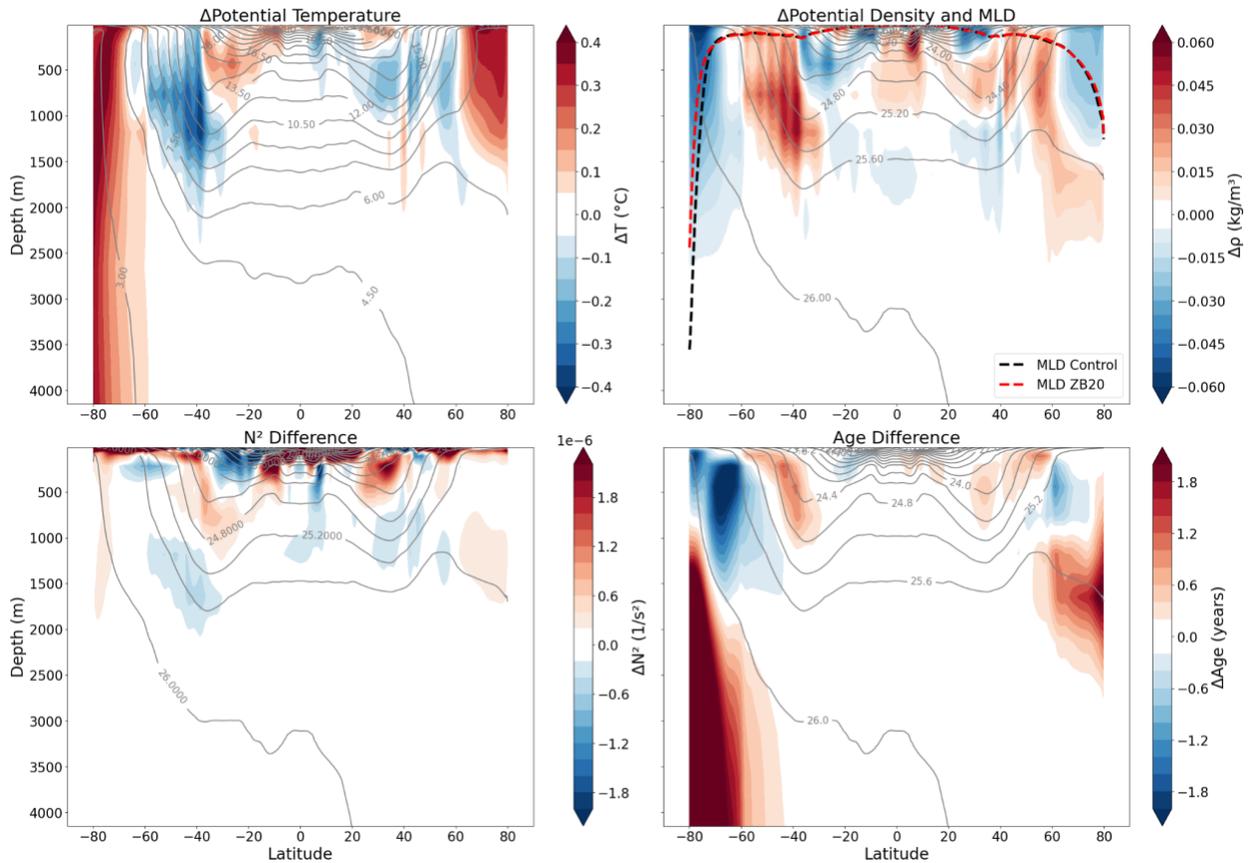

**Figure 6.** Zonal mean pattern of subsurface ocean responses averaged over 40 years. Panels show: (a) subsurface temperature anomalies (ZB20 – Control; shading) with climatological temperature from Control (contours); (b) potential density ($\sigma_0$) anomalies (shading), with winter mixed-layer depths (DJF in the Northern Hemisphere and JJA in the Southern Hemisphere) from the Control and ZB20 shown as dashed black and red curves, respectively; (c) buoyancy frequency $N^2$ anomalies; and (d) water-age anomalies relative to the Control. Contours in panels (b)–(d) indicate the climatological $\sigma_0$ field.

The hemispheric asymmetry in the subsurface response is also clear: the Southern Hemisphere cooling and warming signals are broader and stronger than their Northern Hemisphere counterparts. This contrast arises from the presence of the "Drake Passage", which permits a continuous circumpolar current and enables vigorous baroclinic instability, mesoscale stirring, deep-water formation and meridional heat redistribution. In the absence of land barriers, the Southern Ocean forms a dynamically connected system, allowing the backscatter-induced eddy enhancement to influence both horizontal and vertical heat redistribution. By contrast, the Northern Hemisphere has enclosed basins that constrain eddy-mediated pathways of heat redistribution, resulting in a more localized and weaker response. Furthermore, the Antarctic Circumpolar Current (ACC) transport weakens under the ZB20 parameterization, decreasing from 699 Sv in Control to 645 Sv in ZB20 (a reduction of 54 Sv, around -7.7%). This change indicates an adjustment of the circumpolar circulation in response to the enhanced mesoscale eddy activity. This reduction in ACC transport is dynamically consistent with the concurrent changes in meridional density gradient (Figure 6b) across the ACC based on previous studies (Kuhlbrodt et al. 2012; Shi et al. 2021).

To further examine the dynamical responses, we show the response of the meridional overturning circulation (MOC) in density space (Figure 7). The inclusion of the backscatter scheme leads to an overall enhancement of the overturning strength, most notably in the Southern Hemisphere, where an intensified anticlockwise cell develops and is centered around approximately 60° S. This strengthened circulation reflects the enhanced eddy velocity and the eddy-induced overturning.

In the Northern Hemisphere, the MOC anomaly exhibits a dipolar vertical structure, characterized by a positive anomaly in the upper branch and a negative anomaly below. Relative to the mean state, with a clockwise cell analogous to the Atlantic Meridional Overturning Circulation (AMOC), this dipolar pattern implies an upward shift of the overturning structure, associated with density reduction and thermocline shoaling in the mid- to high-latitude regions. A similar upward displacement is also apparent in the southern subpolar region.

The water age response (Figure 6d) exhibits a meridionally structured dipole that closely mirrors the MOC anomaly (Figure 7b). In the high-latitude Southern Ocean, water age decreases (younger waters), reflecting enhanced ventilation associated with the strengthened overturning cell shown in Figure 7. This younger signal is confined primarily to the upper branch of the circulation, consistent with increased eddy-driven exchange and a more vigorous poleward transport of recently ventilated waters. In contrast, the mid-latitudes display positive water age anomalies (older waters), indicating reduced ventilation where the overturning circulation weakens. Overall, the water age response provides an integrated tracer perspective that is dynamically consistent with the diagnosed MOC adjustments.

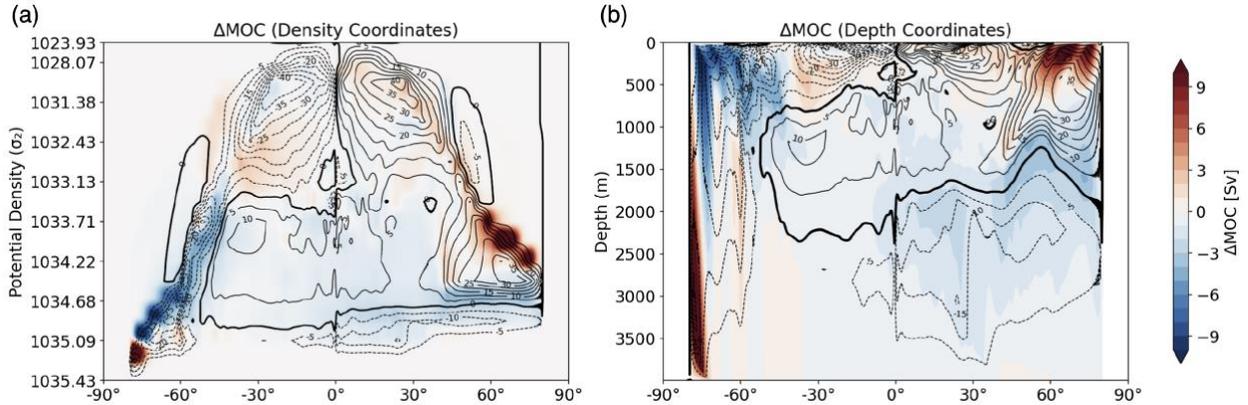

**Figure 7.** Meridional overturning circulation (MOC) response in density coordinates averaged over 40 years. Panel (a) shows the MOC anomalies in $\sigma_2$ density space, and panel (b) shows the same anomalies which are remapped to depth space. The thick contour denotes the zero value of streamfunction, with thin clockwise (positive) and anticlockwise (negative) circulation, shown by solid (dashed) contours, respectively.

### 3.5 Responses of Meridional Heat Transport in Ocean and Atmosphere

Figure 8 shows the responses of meridional heat transport in both the ocean and the atmosphere under the backscatter parameterization. The oceanic heat transport (OHT) is enhanced from the mid-latitudes to the high latitudes in both hemispheres, with the substantial increase to the south of 40°S. This pattern indicates that the backscatter scheme strengthens poleward ocean heat redistribution, consistent with the subsurface cooling in the thermocline and the subpolar deep warming described earlier. In contrast, the atmospheric heat transport (AHT) exhibits a compensatory decrease over the same latitude bands. This reduction underscores the tight coupling between oceanic and atmospheric processes: as the ocean becomes more efficient in transporting heat poleward, the atmosphere correspondingly reduces its contribution. Such compensation is consistent with the coupled energy-balance constraint that the total meridional heat transport is largely set by the equator-to-pole gradient in solar radiation (e.g. Enderton and Marshall 2009).

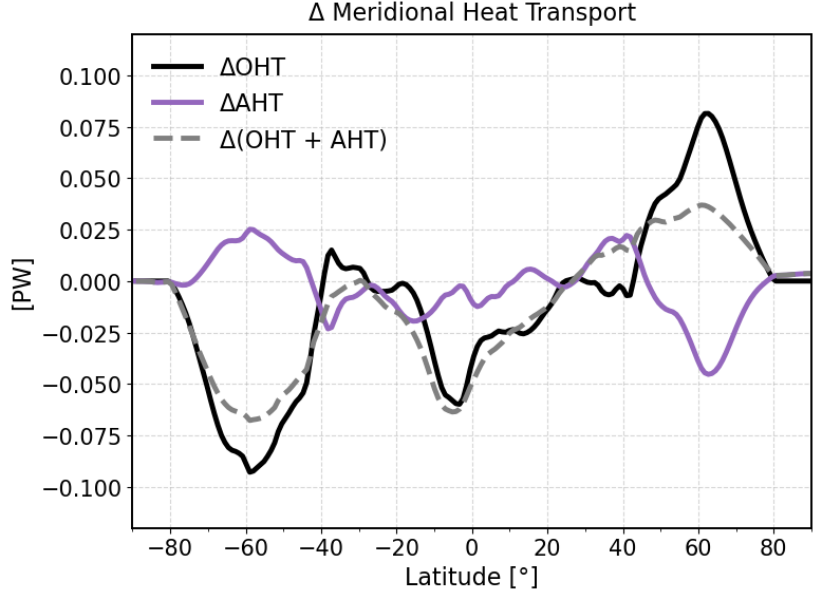

**Figure 8.** Meridional heat transport responses in the ocean (ΔOHT; black curve) and atmosphere (ΔAHT; purple curve) averaged over 40 years. The sum of them is shown as a dashed gray curve.

Figure 9 further decomposes the OHT and AHT responses into contributions from the mean circulation and eddy-driven components. For OHT, the time-mean component is estimated from the product of the time-mean meridional velocity and temperature, and the residual relative to the total OHT represents the transient eddy contribution. The AHT is decomposed into components associated with the meridional overturning circulation (MOC), standing eddies, and transient eddies (Donohoe et al. 2020).

For the OHT response (Figure 9c), the eddy-driven contribution exhibits a distinct, spike-like enhancement centered around 40°–45°S, coinciding with regions of strong mesoscale eddy activity and vigorous baroclinic instability. This localized intensification reflects the enhanced eddy heat flux associated with the backscatter-induced energization of transient eddies (see Figure 2). In contrast, the broader enhancement at higher latitudes arises primarily from the mean circulation component, which strengthens the poleward advection of heat into the subpolar regions.

For the AHT response (Figure 9d), transient eddies play the dominant role in reducing poleward heat transport. Relative to the climatology of AHT (Figure 9b), the anomaly exhibits a pronounced equatorward shift, mirroring the mid-latitude wind stress response (Figure 5). This shift suggests a displacement of the baroclinic zones and storm tracks toward the equator, linked to the altered meridional temperature gradient induced by the oceanic changes.

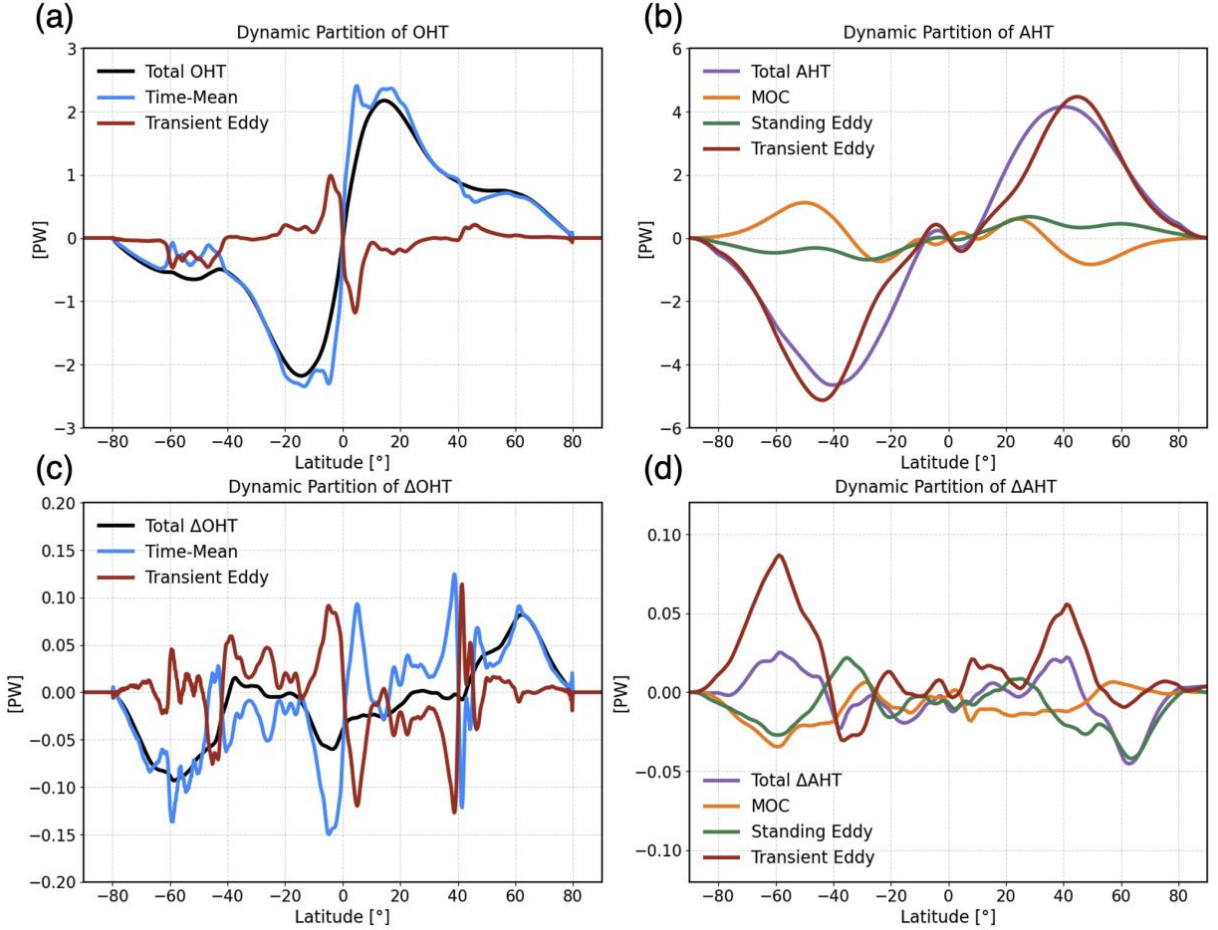

**Figure 9.** Decomposition of ocean heat transport (OHT) and atmospheric heat transport (AHT) and their responses (ZB20 - Control) averaged for 40 years. Panels (a) and (b) show the climatological decomposition of OHT and AHT from Control, respectively, and panels (c) and (d) show the responses of each component. For OHT, the transport is decomposed into a time-mean component (blue) and a transient-eddy component (red). For AHT, the transport is decomposed into contributions from the meridional overturning circulation (MOC; orange), standing eddies (green), and transient eddies (brown).

To further elucidate the mechanisms linking the mid-latitude cooling and high-latitude warming seen in Figure 6, a regional heat budget analysis was performed using two latitude bands: 40°S–60°S and 60°S–80°S (Figure 10). The analysis quantifies the relative roles of meridional heat transport convergence and net surface heat flux anomalies in driving local changes in ocean heat content (OHC).

Results indicate that meridional heat transport convergence dominates the OHC tendency and explains the observed dipolar temperature pattern between mid- and high latitudes (Figure 10b).

In both regions, the contribution from net surface heat flux anomalies is small (Figure 10c), implying that the SST and subsurface temperature changes primarily arise from redistribution of oceanic heat rather than direct surface forcing. Within the 40°S–60°S band, the mean and eddy components of heat transport are comparable in magnitude, jointly contributing to the local cooling and reduced OHC (Figure 10b). In contrast, in the 60°S–80°S band, the heat convergence and resulting subsurface warming are dominated by the mean circulation.

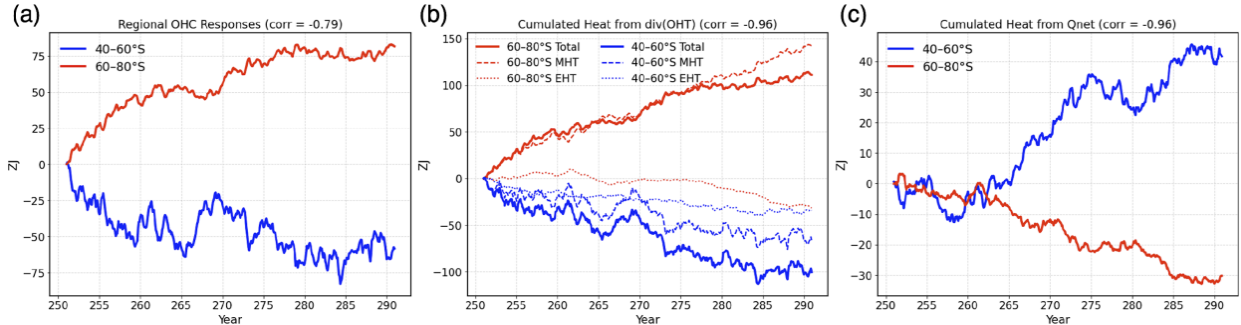

**Figure 10.** Regional heat budget for 40°S–60°S (blue) and 60°S–80°S (red) for the 40 years of simulation. Panels show (a) ocean heat content (OHC) anomalies from ZB20 relative to Control in the two latitude bands, (b) contributions from meridional heat transport divergence (i.e. div(OHT)), and (c) contributions from net surface heat flux anomalies (i.e. Qnet). In panel (b), the contributions from the mean heat transport (MHT) and eddy heat transport (EHT) are shown as dashed and dotted curves, respectively. The correlation coefficient between the accumulated OHC anomalies in the two latitude bands is shown in brackets.

## 4. Sensitivity to Parameterization Scaling Coefficient

To assess the robustness of the data-driven eddy backscatter scheme and quantify its dynamical influence, a series of sensitivity experiments were conducted by varying the scaling coefficient $\gamma$ from 1.5 to 3.0. The responses of several key dynamical quantities, including kinetic energy, eddy activity, meridional overturning, and regional heat content, exhibit a broadly linear dependence on $\gamma$ (Figure 11), indicating that these responses arise from the parameterization rather than random noise.

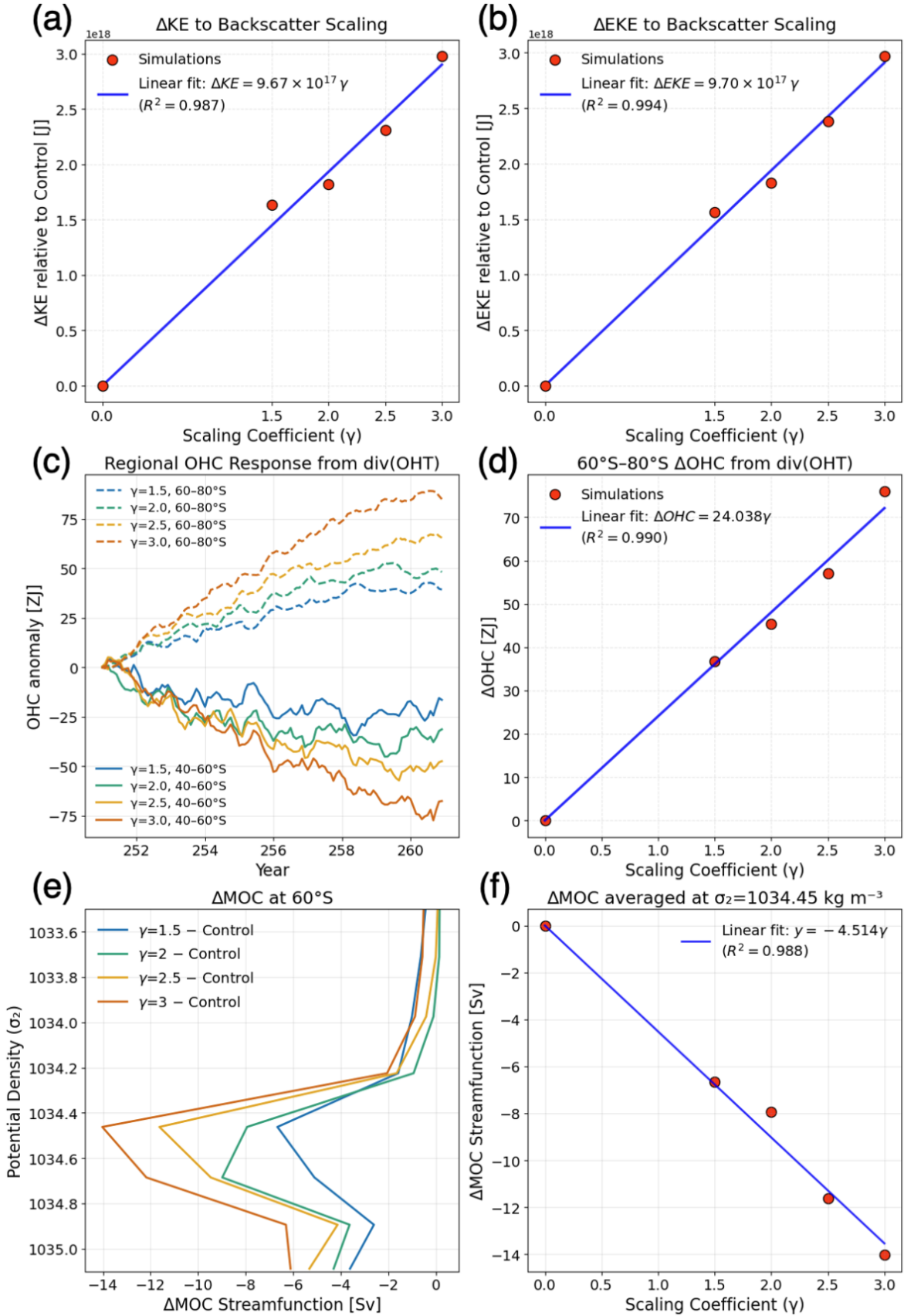

**Figure 11.** Sensitivity of the responses to the scaling coefficient γ used in the eddy parameterization for the first 10 years of simulation. Panels show (a) total KE anomalies, (b) EKE anomalies, (c) time evolution of accumulated OHC anomalies in the 40°S–60°S and 60°S–80°S latitude bands due to meridional heat transport divergence, (d) OHT divergence contributed OHC anomalies within 60°S–80°S, and (e–f) MOC strength anomalies at 60˚S and its averaged value at 1034.45 kg m$^{-3}$ density level. The points at γ=0 correspond to the Control run. All anomalies are relative to the Control run. Linear fits are of the form $y = \alpha\gamma$, and the R² values are shown in brackets.

The total KE increases almost linearly with γ, with a best-fit slope of approximately $O(10^{18})$ J per unit γ (Figure 11a). The EKE shows an even clearer linear trend, reinforcing the interpretation that the parameterization primarily energizes transient mesoscale eddies (Figure 11b). Moreover, the accumulated heat anomalies associated with the divergence of OHT in both mid- and high-latitude bands scale with γ, demonstrating that stronger backscatter consistently enhances poleward heat redistribution (Figures 11c,d). In the 40°S–60°S band, this manifests as progressively stronger heat-content reductions, whereas in the 60°S–80°S region, OHC increases with γ.

The meridional overturning circulation (MOC) exhibits a similarly linear response to γ. The enhancement of the overturning near ~60°S strengthens with increasing γ (Figure 11e). At $\sigma_2 = 1034.45$ kg m$^{-3}$ density level, which has the largest response, the MOC anomaly increases approximately linearly with γ (Figure 11f), indicating that backscatter modulates the large-scale circulation in a predictable manner.

Collectively, these results indicate that the data-driven eddy backscatter parameterization behaves in a stable and well-conditioned manner across a wide range of amplitudes. Importantly, the absence of nonlinear amplification or runaway behavior implies that the parameterization remains dynamically controlled even at the upper bound of γ considered here. This predictable scaling provides confidence for applications in more realistic model configurations.

## 5. Conclusions and Discussion

This study investigates the impact of a data-driven eddy momentum parameterization, namely the Zanna and Bolton (2020) equation-discovery closure, on the coupled ocean–atmosphere climate system using an idealized global model configuration. The numerical experiments highlight several key findings.

- The implementation of the eddy momentum parameterization substantially enhances mesoscale eddy activity, as reflected in the pronounced increase in eddy kinetic energy, particularly in the mid-latitudes. The effect of the momentum parameterization is to

- backscatter energy to larger scales, thereby effectively energizing dynamically-active regions, strengthening barotropic and baroclinic instabilities without generating artificial eddies elsewhere.
- The intensified eddy activity strengthens the oceanic meridional heat transport, particularly from the mid-latitudes toward the subpolar regions, with a ~20% increase at 60°S for a scaling coefficient of 2.0. This enhanced poleward heat redistribution leads to a characteristic dipolar temperature pattern of surface and subsurface cooling in the mid-latitudes and warming at high latitudes. The associated changes in ocean heat content are primarily driven by ocean meridional heat transport convergence rather than surface heat flux anomalies.
- The atmospheric meridional heat transport acts to compensate for enhanced ocean heat transport by transporting less heat poleward, maintaining the total (ocean + atmosphere) heat transport constrained by the equator-to-pole gradient set by the incoming solar radiation (Enderton and Marshall 2009).
- The Southern Hemisphere shows a stronger and broader response than the Northern Hemisphere, owing to the existence of an open circumpolar channel analogous to the Drake Passage and Southern Ocean Circulation. The open circumpolar channel permits a continuous flow that supports sustained eddy activity, thereby enhancing eddy stirring and heat redistribution, and contributing to the pronounced interhemispheric asymmetry in this study.
- Finally, the zonal-mean wind system shifts equatorward, associated with the mid-latitude oceanic cooling and reduced meridional temperature gradient. This shift implies a concurrent displacement of the baroclinic zones and storm tracks, linking oceanic dynamical changes to large-scale atmospheric circulation responses.

Despite its idealized design, this study provides a clear dynamical interpretation of how a data-driven eddy backscatter scheme can modulate the coupled climate state. Several points merit further discussion.

The model employs an idealized bathymetry and thus lacks a direct observational target for quantitative validation. Moreover, running higher-resolution coupled simulations which explicitly resolve eddy would be computationally prohibitive, even in an idealized configuration. Consequently, our analysis focuses on the initial sensitivity of the coupled model to the ZB20 backscatter parameterization (Zanna and Bolton 2020) and the associated physical processes, rather than on absolute magnitudes or equilibrium responses. This approach provides valuable guidance for improving parameterizations in more realistic climate models. For example, in previous studies using comprehensive coupled models, the Southern Ocean often exhibits a warm bias to the south of 40˚S relative to observations (Q. Zhang et al. 2023; Luo et al. 2023). In particular, Juricke et al. (2020) examined this bias using an ocean–sea ice model that incorporates eddy backscatter effects. In our simulations, the ZB20 backscatter scheme induces mid--latitude cooling in the Southern Hemisphere, suggesting its potential to alleviate the warm bias, which is consistent with Juricke et al. (2020).

Spectral analyses of EKE show that while the eddy energy increases, there is no substantial shift in eddy length scale. This indicates that the backscatter primarily enhances the intensity of mesoscale motions rather than altering their scale. According to mixing length theory (e.g. Prandtl 1925; Kong and Jansen 2021), the strengthened eddy field promotes isopycnal mixing, facilitating enhanced heat exchange between mid- and high-latitude regions, consistent with the OHT and temperature responses. For instance, increased isopycnal mixing has been shown to cool the Southern Ocean (e.g. Holmes et al. 2022; Neumann and Jones 2025).

Although the model includes an active sea-ice model, no sea ice forms under the climate conditions, similar to Wu et al. (2021). Therefore, sea-ice effects are not involved in the responses shown here, and the feedbacks arise primarily from air–sea coupling. In a more realistic case with sea ice formation, the parameterization-induced polar warming would potentially reduce ice coverage. Considering the associated albedo feedback (e.g. Holland and Bitz 2003, Chung et al. 2025), this may further enhance high-latitude warming. However, this effect may also depend on the background ice state and the coupled ocean–ice–atmosphere balance, highlighting the need for future studies in a more realistic sea-ice regime.

The simulated equatorward shift of the westerly jet is primarily ocean-driven, due to altered SST gradients. However, the atmospheric response can feed back onto the ocean through anomalous Ekman transport and upwelling, modulating the surface heat flux and subsurface temperature structure. The current model captures this coupled feedback implicitly, but a detailed attribution separating ocean-forced and atmosphere-driven components would require additional targeted experiments, such as partially coupled configurations (Luongo et al. 2024; McMonigal et al. 2025).

Moreover, the Hadley circulation appears largely unaffected (not shown), and the cross-equatorial heat transport shows minimal change. This weak response is likely related to the hemispherically symmetric idealized geometry and similar ocean basin areas, which constrain interhemispheric energy exchange in the atmosphere. The dominant effects thus remain concentrated within each hemisphere's mid- to high-latitude regimes.

This study has focused on the mean-state adjustments induced by the data-driven eddy parameterization. An important next step is to assess how such schemes influence modes of climate variability, such as ENSO-like oscillations and decadal variability, which may respond differently than the mean state. Clarifying these impacts, in more realistic simulations, will be essential for understanding the broader implications of mesoscale eddy parameterization in the fully coupled climate system.

**Acknowledgments**

This project is supported by Schmidt Sciences through the M$^2$LInES project. This research used resources of the National Energy Research Scientific Computing Center (NERSC), a U.S. Department of Energy Office of Science User Facility. We also acknowledge high-performance computing support from the Derecho system provided by the NSF National Center for Atmospheric Research (NCAR), sponsored by the National Science Foundation. We also thank D. Balwada, F. Lu, M. Bushuk, B. Reichl, M. Pudig for their invaluable feedback on this work.

**Data Availability Statement**

The CESM source code is available at https://www.cesm.ucar.edu/models/simple. The MOM6 source code with ZB20 parameterization is available at https://github.com/mom-ocean/MOM6.

# References

Abernathey, R., D. Ferreira, and A. Klocker, 2013: Diagnostics of isopycnal mixing in a circumpolar channel. *Ocean Model.*, **72**, 1–16, https://doi.org/10.1016/j.ocemod.2013.07.004.

Adcroft, A., and Coauthors, 2019: The GFDL Global Ocean and Sea Ice Model OM4.0: Model Description and Simulation Features. *J. Adv. Model. Earth Syst.*, **11**, 3167–3211, https://doi.org/10.1029/2019MS001726.

Bailey, D., A. DuVivier, M. Holland, E. Hunke, B. Lipscomb, B. Briegleb, C. Bitz, and J. Schramm, 2018: *CESM CICE5 users guide (Tech. Rep.)*, https://app.readthedocs.org/projects/cesmcice/downloads/pdf/latest/.

Beech, N., T. Rackow, T. Semmler, S. Danilov, Q. Wang, and T. Jung, 2022: Long-term evolution of ocean eddy activity in a warming world. *Nat. Clim. Chang.*, 12, 910–917, https://doi.org/10.1038/s41558-022-01478-3.

Bian, C., Z. Jing, H. Wang, L. Wu, Z. Chen, B. Gan, and H. Yang, 2023: Oceanic mesoscale eddies as crucial drivers of global marine heatwaves. *Nat Commun*, 14, 2970, https://doi.org/10.1038/s41467-023-38811-z.

Chelton, D. B., R. A. deSzoeke, M. G. Schlax, K. El Naggar, and N. Siwertz, 1998: Geographical Variability of the First Baroclinic Rossby Radius of Deformation. *J. Phys. Oceanogr.*, **28**, 433–460, https://doi.org/10.1175/1520-0485(1998)028%253C0433:GVOTFB%253E2.0.CO;2.

Christensen, H., and L. Zanna, 2022: Parametrization in Weather and Climate Models. *Oxford Research Encyclopedia of Climate Science*, Oxford University Press, https://doi.org/10.1093/acrefore/9780190228620.013.826.

Chung, E.-S., S.-J. Kim, K.-J. Ha, M. F. Stuecker, S.-S. Lee, J.-H. Kim, S.-Y. Jun, and T. Bódai, 2025: The role of sea ice in present and future Arctic amplification. *Commun Earth Environ*, **6**, 910, https://doi.org/10.1038/s43247-025-02834-9.

Danabasoglu, G., and Coauthors, 2020: The Community Earth System Model Version 2 (CESM2). *J. Adv. Model. Earth Syst.*, **12**, 1–35, https://doi.org/10.1029/2019MS001916.

Delworth, T. L., and Coauthors, 2012: Simulated climate and climate change in the GFDL CM2.5 high-resolution coupled climate model. *J. Clim.*, **25**, 2755–2781, https://doi.org/10.1175/JCLI-D-11-00316.1.

Donohoe, A., K. C. Armour, G. H. Roe, D. S. Battisti, and L. Hahn, 2020: The Partitioning of Meridional Heat Transport from the Last Glacial Maximum to CO2 Quadrupling in Coupled Climate Models. *J. Clim.*, **33**, 4141–4165, https://doi.org/10.1175/jcli-d-19-0797.1.

Enderton, D., and J. Marshall, 2009: Explorations of Atmosphere–Ocean–Ice Climates on an Aquaplanet and Their Meridional Energy Transports. *J. Atmospheric Sci.*, **66**, 1593–1611, https://doi.org/10.1175/2008JAS2680.1.

Eyring, V., S. Bony, G. A. Meehl, C. A. Senior, B. Stevens, R. J. Stouffer, and K. E. Taylor, 2016: Overview of the Coupled Model Intercomparison Project Phase 6 (CMIP6) experimental design and organization. *Geosci. Model Dev.*, **9**, 1937–1958, https://doi.org/10.5194/gmd-9-1937-2016.

Farneti, R., and G. K. Vallis, 2009: An Intermediate Complexity Climate Model (ICCMp1) based on the GFDL flexible modelling system. *Geosci. Model Dev.*, **2**, 73–88, https://doi.org/10.5194/gmd-2-73-2009.

Ferrari, R., and C. Wunsch, 2009: Ocean Circulation Kinetic Energy: Reservoirs, Sources, and Sinks. *Annu. Rev. Fluid Mech.*, **41**, 253–282, https://doi.org/10.1146/annurev.fluid.40.111406.102139.

Frezat, H., J. Le Sommer, R. Fablet, G. Balarac, and R. Lguensat, 2022: A Posteriori Learning for Quasi-Geostrophic Turbulence Parametrization. *J. Adv. Model. Earth Syst.*, **14**, e2022MS003124, https://doi.org/10.1029/2022MS003124.

Gent, P. R., and J. C. Mcwilliams, 1990: Isopycnal Mixing in Ocean Circulation Models. *J. Phys. Oceanogr.*, **20**, 150–155, https://doi.org/10.1175/1520-0485(1990)020%253C0150:IMIOCM%253E2.0.CO;2.

Germano, M., 1986. A proposal for a redefinition of the turbulent stresses in the filtered Navier–Stokes equations. *The Physics of Fluids*, **29**(7), 2323–2324. https://doi.org/10.1063/1.865568.

Griffies, S. M., and R. W. Hallberg, 2000: Biharmonic Friction with a Smagorinsky-Like Viscosity for Use in Large-Scale Eddy-Permitting Ocean Models. *Mon. Weather Rev.*, **128**, 2935–2946, https://doi.org/10.1175/1520-0493(2000)128%253C2935:BFWASL%253E2.0.CO;2.

——, and Coauthors, 2015: Impacts on Ocean Heat from Transient Mesoscale Eddies in a Hierarchy of Climate Models. *J. Clim.*, **28**, 952–977, https://doi.org/10.1175/JCLI-D-14-00353.1.

Guan, Y., A. Chattopadhyay, A. Subel, and P. Hassanzadeh, 2022: Stable a posteriori LES of 2D turbulence using convolutional neural networks: Backscattering analysis and generalization to higher Re via transfer learning. *J. Comput. Phys.*, **458**, 111090, https://doi.org/10.1016/j.jcp.2022.111090.

Guillaumin, A. P., and L. Zanna, 2021: Stochastic-Deep Learning Parameterization of Ocean Momentum Forcing. *J. Adv. Model. Earth Syst.*, **13**, e2021MS002534, https://doi.org/10.1029/2021MS002534.

Hallberg, R., 2013: Using a resolution function to regulate parameterizations of oceanic mesoscale eddy effects. *Ocean Model.*, **72**, 92–103, https://doi.org/10.1016/j.ocemod.2013.08.007.

Hewitt, H. T., and Coauthors, 2020: Resolving and Parameterising the Ocean Mesoscale in Earth System Models. *Curr. Clim. Change Rep.*, **6**, 137–152, https://doi.org/10.1007/s40641-020-00164-w.

Holland, M. M., and C. M. Bitz, 2003: Polar amplification of climate change in coupled models. *Clim. Dyn.*, **21**, 221–232, https://doi.org/10.1007/s00382-003-0332-6.

Holmes, R. M., S. Groeskamp, K. D. Stewart, and T. J. McDougall, 2022: Sensitivity of a Coarse-Resolution Global Ocean Model to a Spatially Variable Neutral Diffusivity. *J. Adv. Model. Earth Syst.*, **14**, https://doi.org/10.1029/2021ms002914.

Hurrell, J. W., and Coauthors, 2013: The community earth system model: A framework for collaborative research. *Bull. Am. Meteorol. Soc.*, **94**, 1339–1360, https://doi.org/10.1175/BAMS-D-12-00121.1.

Jansen, M. F., and I. M. Held, 2014: Parameterizing subgrid-scale eddy effects using energetically consistent backscatter. *Ocean Model.*, **80**, 36–48, https://doi.org/10.1016/j.ocemod.2014.06.002.

——, ——, A. Adcroft, and R. Hallberg, 2015: Energy budget-based backscatter in an eddy permitting primitive equation model. *Ocean Model.*, **94**, 15–26, https://doi.org/10.1016/j.ocemod.2015.07.015.

Juricke, S., S. Danilov, A. Kutsenko, and M. Oliver, 2019: Ocean kinetic energy backscatter parametrizations on unstructured grids: Impact on mesoscale turbulence in a channel. *Ocean Model.*, **138**, 51–67, https://doi.org/10.1016/j.ocemod.2019.03.009.

Juricke, S., S. Danilov, N. Koldunov, M. Oliver, and D. Sidorenko, 2020: Ocean Kinetic Energy Backscatter Parametrization on Unstructured Grids: Impact on Global Eddy-Permitting Simulations. Journal of Advances in Modeling Earth Systems, 12, https://doi.org/10.1029/2019MS001855.

Klöwer, M., M. F. Jansen, M. Claus, R. J. Greatbatch, and S. Thomsen, 2018: Energy budget-based backscatter in a shallow water model of a double gyre basin. *Ocean Model.*, **132**, 1–11, https://doi.org/10.1016/j.ocemod.2018.09.006.

Kong, H., and M. F. Jansen, 2021: The Impact of Topography and Eddy Parameterization on the Simulated Southern Ocean Circulation Response to Changes in Surface Wind Stress. Journal of Physical Oceanography, 51, 825–843, https://doi.org/10.1175/JPO-D-20-0142.1.

Kuhlbrodt, T., R. S. Smith, Z. Wang, and J. M. Gregory, 2012: The influence of eddy parameterizations on the transport of the Antarctic Circumpolar Current in coupled climate models. Ocean Modelling, 52–53, 1–8, https://doi.org/10.1016/j.ocemod.2012.04.006.

Lawrence, D. M., and Coauthors, 2019: The Community Land Model Version 5: Description of New Features, Benchmarking, and Impact of Forcing Uncertainty. *J. Adv. Model. Earth Syst.*, **11**, 4245–4287, https://doi.org/10.1029/2018MS001583.

Luo, F., J. Ying, T. Liu, and D. Chen, 2023: Origins of Southern Ocean warm sea surface temperature bias in CMIP6 models. *Npj Clim. Atmospheric Sci.*, **6**, 127, https://doi.org/10.1038/s41612-023-00456-6.

Luongo, M. T., N. G. Brizuela, I. Eisenman, and S. Xie, 2024: Retaining Short-Term Variability Reduces Mean State Biases in Wind Stress Overriding Simulations. *J. Adv. Model. Earth Syst.*, **16**, e2023MS003665, https://doi.org/10.1029/2023MS003665.

McMonigal, K., S. M. Larson, and M. Gervais, 2025: Wind-Driven Ocean Circulation Changes Can Amplify Future Cooling of the North Atlantic Warming Hole. *J. Clim.*, **38**, 2479–2496, https://doi.org/10.1175/JCLI-D-24-0227.1.

Neale, R. B., J. Richter, S. Park, P. H. Lauritzen, S. J. Vavrus, P. J. Rasch, and M. Zhang, 2013: The Mean Climate of the Community Atmosphere Model (CAM4) in Forced SST and Fully Coupled Experiments. *J. Clim.*, **26**, 5150–5168, https://doi.org/10.1175/JCLI-D-12-00236.1.

Neumann, N. K., and C. S. Jones, 2025: Effects of Wind and Isopycnal Mixing on Southern Ocean Surface Buoyancy Flux and Antarctic Bottom Water Formation. *Geophys. Res. Lett.*, **52**, https://doi.org/10.1029/2024gl112133.

Perezhogin, P, and A. Glazunov, 2023: Subgrid parameterizations of ocean mesoscale eddies based on Germano decomposition. *J. Adv. Model. Earth Syst,* 15(10), e2023MS003771, https://doi.org/10.1029/2023MS003771.

Perezhogin, P., C. Zhang, A. Adcroft, C. Fernandez-Granda, and L. Zanna, 2024: A Stable Implementation of a Data-Driven Scale-Aware Mesoscale Parameterization. *J. Adv. Model. Earth Syst.*, **16**, e2023MS004104, https://doi.org/10.1029/2023MS004104.

Prandtl, L., 1925: Bericht über Untersuchungen zur ausgebildeten Turbulenz. Z Angew Math Mech, 5, 136–139, https://doi.org/10.1002/zamm.19250050212.

Redi, M. H., 1982: Oceanic Isopycnal Mixing by Coordinate Rotation. *J. Phys. Oceanogr.*, **12**, 1154–1158, https://doi.org/10.1175/1520-0485(1982)012%253C1154:OIMBCR%253E2.0.CO;2.

Rintoul, S. R., 2018: The global influence of localized dynamics in the Southern Ocean. *Nature*, **558**, 209–218, https://doi.org/10.1038/s41586-018-0182-3.

Shi, J.-R., L. D. Talley, S.-P. Xie, Q. Peng, and W. Liu, 2021: Ocean warming and accelerating Southern Ocean zonal flow. *Nature Climate Change*, 11, 1090–1097, https://doi.org/10.1038/s41558-021-01212-5.

Smith, R. S., C. Dubois, and J. Marotzke, 2006: Global Climate and Ocean Circulation on an Aquaplanet Ocean–Atmosphere General Circulation Model. *J. Clim.*, **19**, 4719–4737, https://doi.org/10.1175/JCLI3874.1.

Stephens, G. L., and Coauthors, 2012: An update on Earth's energy balance in light of the latest global observations. *Nat. Geosci.*, **5**, 691–696, https://doi.org/10.1038/ngeo1580.

Trenberth, K. E., and J. M. Caron, 2001: Estimates of Meridional Atmosphere and Ocean Heat Transports. *J. Clim.*, **14**, 3433–3443, https://doi.org/10.1175/1520-0442(2001)014%253C3433:EOMAAO%253E2.0.CO;2.

Uchida, T., R. Abernathey, and S. Smith, 2017: Seasonality of eddy kinetic energy in an eddy permitting global climate model. *Ocean Model.*, **118**, 41–58, https://doi.org/10.1016/j.ocemod.2017.08.006.

Wolfe, C. L., and P. Cessi, 2010: What Sets the Strength of the Middepth Stratification and Overturning Circulation in Eddying Ocean Models? *J. Phys. Oceanogr.*, **40**, 1520–1538, https://doi.org/10.1175/2010JPO4393.1.

Wu, X., K. A. Reed, C. L. P. Wolfe, G. M. Marques, S. D. Bachman, and F. O. Bryan, 2021: Coupled Aqua and Ridge Planets in the Community Earth System Model. J Adv Model Earth Syst, 13, e2020MS002418, https://doi.org/10.1029/2020MS002418.

Wunsch, C., 2005: The Total Meridional Heat Flux and Its Oceanic and Atmospheric Partition. *J. Clim.*, **18**, 4374–4380, https://doi.org/10.1175/JCLI3539.1.

Zanna, L., and T. Bolton, 2020: Data-Driven Equation Discovery of Ocean Mesoscale Closures. *Geophys. Res. Lett.*, **47**, e2020GL088376, https://doi.org/10.1029/2020GL088376.

Zhang, C., P. Perezhogin, C. Gultekin, A. Adcroft, C. Fernandez-Granda, and L. Zanna, 2023: Implementation and Evaluation of a Machine Learned Mesoscale Eddy Parameterization Into a Numerical Ocean Circulation Model. *J. Adv. Model. Earth Syst.*, **15**, e2023MS003697, https://doi.org/10.1029/2023MS003697.

Zhang, Q., B. Liu, S. Li, and T. Zhou, 2023: Understanding Models' Global Sea Surface Temperature Bias in Mean State: From CMIP5 to CMIP6. *Geophys. Res. Lett.*, **50**, e2022GL100888, https://doi.org/10.1029/2022GL100888.